\documentclass[aip,twocolumn,showpacs,amsmath,amssymb,jap,reprint]{revtex4-1}
\usepackage[dvipdfmx]{graphicx}%
\usepackage{dcolumn}
\usepackage{bm}
\usepackage{braket}

\def\E{{\mathcal{E}}}
\def\T{{\mathcal{T}}}
\begin{document}
\title{Macroscopic electron-hole distribution in silicon and cubic silicon carbide}
\author{T. Otobe}
\affiliation{Kansai Photon Science Institute, National Institutes for Quantum and Radiological Science and Technology (QST), Kyoto 619-0215, Japan}
\begin{abstract}
Electron excitations at silicon and 3C-SiC surfaces caused by an intense femtosecond laser pulse can be calculated
by solving the time-dependent density functional theory and the Maxwell's equation simultaneously.
The energy absorption, carrier density, and electron-hole quasi-temperatures decrease exponentially in 100 nm from the surface.
The electron and hole quasi-temperatures have finite values even at large distances from the surface because of a specific photo-absorption channel.
Although the quasi-temperature in the silicone shows smooth exponential descrease, 3C-SiC shows stepwise decrease because of the change of concerning
bands. 
The quasi-temperature depends not only on the excitation process, i.e., tunnel and multi-photon absorption, but also on the band structure significantly. 
\end{abstract}
\maketitle
\section{introduction}
Processing of solid materials using femtosecond laser pulses
has attracted considerable interest for potential application
to high-precision processing technology. 
\cite{Chichkov,Stuart,Liu,Lenzner,Geissler,Lenzner00,Sudrie,Doumy,Amoruso,Gattass08, Gamaly11,Chimier} 
Because a femtosecond
laser pulse can deposit large amounts of energy into solid
materials within a much shorter time than the conventional spatial diffusion of
thermal energy to the exterior of the irradiated spot, we can
process materials with small thermal denaturation outside of the
irradiated volume.\cite{Gattass08,  Gamaly11}

The precise description of the electron-hole distribution and their quasi-temperatures are important to understand the initial stage of laser processing.
The multi-photon absorption and tunnel excitation processes are the crucial electron excitation processes in semiconductors under the femtosecond laser pulse.
Since these processes are nonlinear and/or nonperturbative, we have to treat the dynamics of the electron and electromagnetic fields simultaneously \cite{Li06,reth17}.

The two temperature model \cite{anishimov75,allen87} (TTM) is a common approach to describe the energy flow between an electron and phonon in metals.
The TTM is applied to semiconductors and dielectrics by combining electron excitation and scattering models\cite{Silaeva_2012,Gallais15}.
Although the electron excitation process and electron-hole distribution is complicated, the excitation process and the estimation of the electron temperature assuming simple model. 
To understand the electron-hole state induced by a laser field around a surface, the dynamics of the electrons and electromagnetic field should be treated directly. 

Recently, we developed a first-principle numerical program SALMON, 
which combines time-dependent density functional theory (TDDFT) and the Maxwell's equation \cite{salmon}.
In the present work, we would like to calculate the macroscopic carrier and hole distribution at the surface of silicon and cubic silicon carbide (3C-SiC).

\section{Computational method}
The theory and its implementation in the present
calculation have been described elsewhere \cite{Bertsch00, yabana12, salmon}; therefore, we describe
it briefly. The laser pulse that enters from the vacuum and
attenuates in the medium varies on a scale of micrometers,
while the electron dynamics take place on a subnanometer
scale. To overcome these conflicting spatial scales, we have
developed a multiscale implementation by introducing two coordinate
systems: macroscopic coordinate $X$ for the laser pulse
propagation and the microscopic coordinate $r$ for local electron
dynamics. The laser pulse is described by the vector potential $\vec{A}_X (t)$, which satisfies
\begin{equation}
\frac{1}{c^2}\frac{\partial^2 \vec{A}_X (t)}{\partial t^2}-\frac{\partial^2 \vec{A}_X (t)}{\partial X^2}=-\frac{4\pi e^2}{c}\vec{J}_X(t).
\label{MXE}
\end{equation}

At each point $X$, we consider lattice-periodic electron dynamics
driven by the electric field $\E_X(t)=-\frac{1}{c}dA_X(t)/dt$. They
are described by the electron orbitals $\psi_{i,X}(\vec{r},t)$, which satisfy
the time-dependent Kohn--Sham equation at each point $X$ \cite{salmon}.
\begin{eqnarray}
i\hbar \frac{\partial}{\partial t} \psi_{i,X}(\vec{r},t)&=&\Bigg[ \frac{1}{2m} \left( -i\hbar\nabla_r+\frac{e}{c}\vec{A}_X(t)\right)^2-\phi_X(\vec{r},t)\nonumber\\
&+&\mu_{xc,X}(\vec{r},t)\Bigg] \psi_{i,X}(\vec{r},t),
\label{TDKS}
\end{eqnarray}
where the potential $\phi_X(\vec{r},t)$, which includes Hartree and
ionic contributions, and the exchange-correlation potential $\mu_{xc,X}(\vec{r},t)$ are periodic in the lattice. 
 In this work, we use local density approximation (LDA) \cite{LDA} for exchange-correlation potential in adiabatic approximation.

The electric current at $X$, $J_X(t)$,
is provided from electron orbitals as
\begin{eqnarray} 
J_X(t) &=& -\frac{e}{mV}  \int_{V} d\vec r \sum_i 
{\rm Re} \psi_{i,X}^* \left( \vec p + \frac{e}{c}\vec A_X(t) \right) \psi_{i,X} \nonumber\\
&+& J_{X,NL}(t), 
\label{CUR} 
\end{eqnarray} 
where $V$ is the volume of a unit cell.  
$J_{X,NL}(t)$ is the current caused by the nonlocality of pseudopotential. 

We solved Eqs.~(\ref{MXE})--(\ref{CUR})
simultaneously as an initial value problem, where the incident
laser pulse is prepared in a vacuum region in front of the
surface, while all the Kohn--Sham orbitals are set to their ground
state.

The incident laser field in vacuum, $\E_{in}(X,t)$,  is described as,
\begin{equation}
\E_{in}(X,t)=
\begin{cases}
\E_0 \sin^2\left(\pi \frac{t_X}{\T_p}\right)\cos(\omega_0 t_X) & 0<t_X<\T_p \\
0& \T_p<t_X< \T_e, \label{eq:field}
\end{cases}
\end{equation} 
where $\E_0$ is the electric field amplitude at the peak,  $\omega_0$ is the laser frequency, $t_X=t-X/c$ describes the space-time dependence of the field.
The pulse length $\T_p$ is set to be 10.81 fs, and the computation is terminated at $\T_e=48.38$ fs.

\section{Results}
\subsection{Silicon}
\begin{figure} 
\includegraphics[width=80mm]{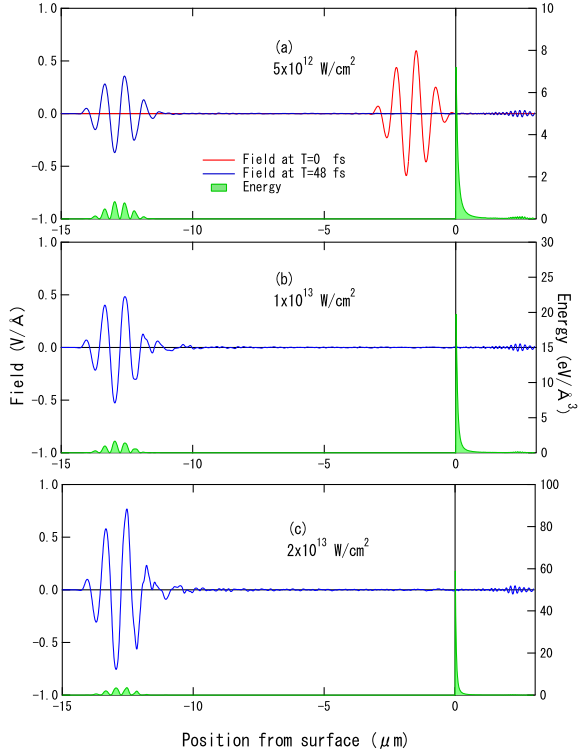} 
\caption{\label{fig1} Energy density and light field after the laser field shine on the silicon surface with 
an intensity of (a) $5\times 10^{12}$, (b) $1\times10^{13}$, and (c) $2\times 10^{13}$ W/cm$^2$.
 The red line represents the initial field (0~fs), and the blue lines represent the field at 48~fs. 
 The energy of densities are represented by the green filled lines. }
\end{figure}
We first assume a typical silicon semiconductor.
Because we use the LDA exchange-correlation potential, the calculated direct band gap (2.4~eV) is smaller than the experimental one (3.1~eV) 
Figure~\ref{fig1} shows the initial laser field (red dashed line in (a)), 
reflected and transmitted field (blue solid lines), and the energy density (green filled lines).
We define the energy of the ground state as 0 eV/\AA$^3$.  

The cubic unit cell containing eight carbon atoms was discretized into grids of 16$^3$. 
The Bloch $k$-space was also discretized into 16$^3$ grid points.
The macroscopic mesh size for the Maxwell Eq. (Eq.~(\ref{MXE})) is 13~nm.
The time step is set to 0.04 atomic unit (0.97 attosecond).

The absorbed energy $E_{ex}$ is defined as the difference of $E_{tot}(t)$ between 
its initial and final values, where $E_{ex}=E_{tot}(\T_e)-E_{tot}(0)$.
The electron-hole density ($N_{\rm e} $) is defined by the projection of the time-dependent wavefunction 
at $t=\T_e$  onto the initial state \cite{Otobe08}.

\begin{figure} 
\includegraphics[width=85mm]{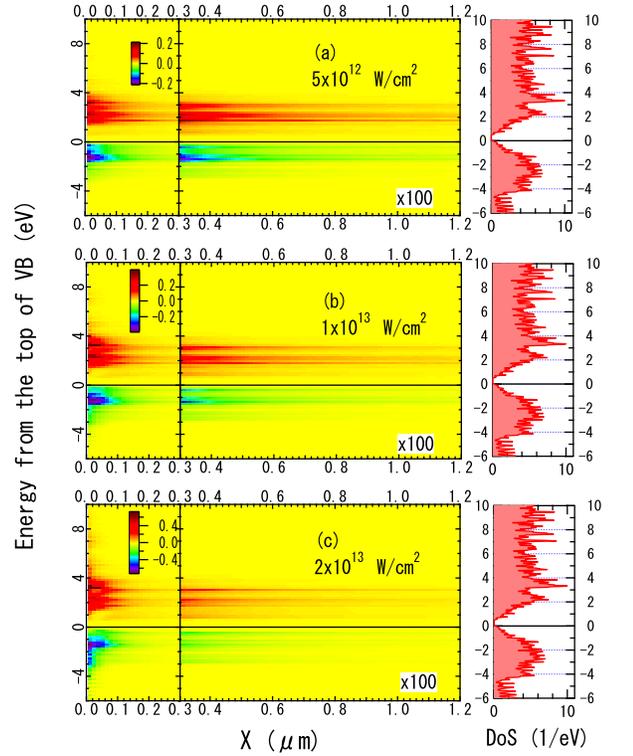} 
\caption{\label{fig2} 
Change of the electron occupation from the initial state. The right panels show the density of states of silicon, for reference.}
 \end{figure}

Individual projections to the initial states at each macro point $X$, given by,  
 \begin{equation}
O^{\vec{k}}_{X,j}(t) = \frac{1}{V}\sum_{i'=occ }  
\vert \langle \Phi^{\vec{k}+\frac{e}{c}\vec{A}(t)}_{X,j} \vert u^{\vec{k}}_{X,i'}(t) \rangle \vert^2 ,
\end{equation}
give the occupation of state $j$. 
We prepare adequate unoccupied states in the conduction band, typically comprising 100 states for each $\vec{k}$, to calculate the overall electron distribution.

Figure~\ref{fig2}  shows the change in the
 electron distribution, 
 \begin{equation}
\delta O^{\vec{k}}_{X,i}=O^{\vec{k}}_{X,i}(\T_e)-O^{\vec{k}}_{X,i}(0),
 \end{equation}
 in the form of  the density of states (DoS).
 Red color indicates the excited electron in the conduction band (CB), whereas the blue indicates the hole in the valence band (VB).
 The color map in the area deeper than 0.3~$\mu m$ is 100 times enhanced. 
 
\begin{figure} 
\includegraphics[width=80mm]{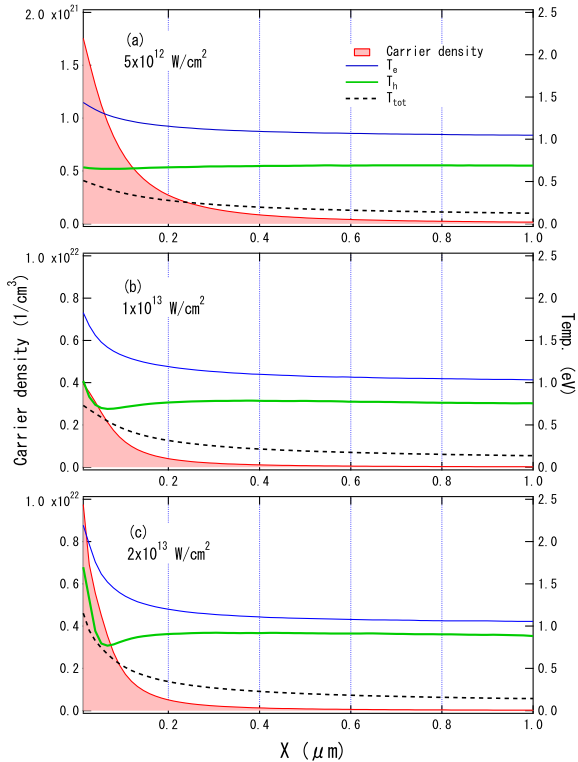} 
\caption{\label{fig3} 
 Position dependence of the carrier density (filled red), electron (blue), and hole (green) temperatures in the quasi-temperature model (QTM).
 The electron temperature in the single-temperature model (STM) is presented by the black dashed lines.}
\end{figure}

  We can estimate the quasi-temperatures of the electron and hole with the QTM, and the temperature of the total system
  with the STM \cite{otobe17}.
The reduced internal energies in the conduction and valence bands, given by $U_{X,c}$ and $U_{X,v}$, respectively, can be defined as 
\begin{equation}
\label{eq:interE}
U_{X,v(c)}=\sum_{\vec{k},i=v(c)} O^{\vec{k}}_{X,i}(\T_e)\epsilon^{\vec{k}}_{X,i} ,
\end{equation}
where $v(c)$ represents the states in the valence (conduction) band, and $\epsilon^{\vec{k}}_{X,i}$ is the energy eigenvalue of the $i$-th state at $X$.
We can assume the quasi-temperatures of carrier ($T_e$) and hole ($T_h$) at each $X$ from $U_{X,v(c)}$, and the temperature ($T_{tot}$) from the $U_{X,tot}=U_{X,v}+U_{X,c}$.

Figure \ref{fig3} shows the carrier density, quasi-temperatures, and temperature as functions of position from the silicon surface.
The color map of the density is 100 times enhanced in regions deeper than $X>0.3~\mu$m.
 At the surface, the electron (red) and hole (blue) spread into a wide energy range as the laser intensity increases.
 However, in the region $X > 0.1~\mu$m, the electron and the hole states have specific peaks whose energy 
 positions do not depend on the position and laser intensity.

The position-independent behavior at $X> 0.1~\mu$m and laser intensity dependence at the surface are presented in Fig.~\ref{fig2},
 and the position-dependent quasi-temperatures are shown in Fig.~\ref{fig3}.
 Around the surface ($X< 0.1 \mu$m), the quasi-temperatures show an exponential decrease as $X$ increases.
 However, in the deeper region, the electron and hole show approximately same quasi-temperatures around 1.0~eV. 
 It should be noted that the carrier density and the electron temperature in the STM show a monotonic decrease, as predicted by the previous models. 
 We can also see a small dip in the hole quasi-temperature around $X=0.05~\mu$m in Fig.~\ref{fig2} (b) and (c), 
 which is the transient region between the exponential decrease and the position-independent region.
 This dip in the hole quasi-temperature corresponds to the intense single peak in hole density.

\begin{figure} 
\includegraphics[width=80mm]{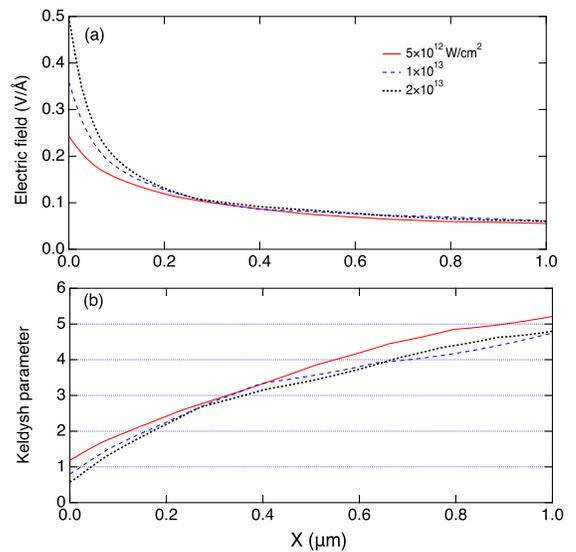} 
\caption{\label{fig4} 
 Position dependence of the (a) maximum field intensity and (b) Keldysh parameter with each initial laser intensities. }
\end{figure}

The dip in the hole quasi-temperature indicates the excitation process dependence in the electron-hole distribution.
In general, the photo-excitation process can be attributed to  regime, that is, multi-photon absorption and tunneling process.
Since an intense laser field induces the tunneling process preferentially, the tunneling process occurs around the surface.
The Keldysh parameter $\gamma=\omega\sqrt{m^* E_{gap}}/e\E$ \cite{Keldysh} is a good parameter to classify the excitation process.
Here, $m^*$ is the effective mass, $E_{gap}$ is the band gap, and $\E$ is the electric field. 
The tunneling process (multi-photon absorption) is dominant for $\gamma \ll 1$ ($\gamma \gg 1$).


Figure \ref{fig4} shows the (a) maximum field and (b) $\gamma$ at $X$.
While the weakest laser intensity ($5\times10^{12}$ W/cm$^2$) shows $\gamma > 1$ for all the $X$ points,
$\gamma$ across the one around $X=0.05$~$\mu$m with a laser intensity of $1\times10^{13}$ and $2\times10^{13}$ W/cm$^2$.
The positions of $\gamma\sim 1$ ($X=0.05$~$\mu$m) correspond to the dip in the hole quasi-temperature, as shown in Fig.~\ref{fig3}. 
This result indicates that the hole and/or electron distribution change by the excitation process.
The tunneling process gives a broader distribution of the electron-hole pairs, which correspond to a higher quasi-temperature.
However, the multiphoton process gives specific peaks of electron-hole pairs, which correspond to the laser intensity--independent 
lower quasi-temperature.

\begin{figure} 
\includegraphics[width=85mm]{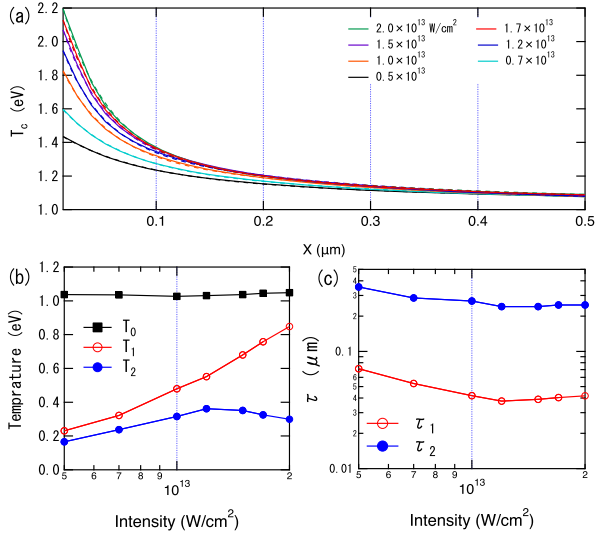} 
\caption{\label{fig5} (a) $T_e(X)$ and the fitted function assuming double exponential (Eq.~(\ref{fit_F})).
(b) Fitting parameters for the temperatures, $T_0$, $T_1$, and $T_2$.
(c) Damping factors for the surface ($\tau_1$) and deeper region ($\tau_2$). }
\end{figure}
Although the hole quasi-temperature shows a complex $X$-dependence, 
the electron quasi-temperature ($T_e$) may be approximated by the simple double exponential function,
\begin{eqnarray}
\label{fit_F}
T_{c}(X)\sim T0+T_1\exp\left[{-\frac{X}{\tau_1}}\right]+T_2\exp\left[-\frac{X}{\tau_2}\right],
\end{eqnarray}
because there are two different excitation processes.
Here, $T_1$ and $\tau_1$ are the peak temperature and the damping factor at the surface, $T_2$ and $\tau_2$ are 
the parameters in a deeper region, $T_0$ is the parameter for saturated temperature.
Figure\ref{fig5} (a) shows the computational results (solid lines) and fitted function (dashed lines).
The fitted function well reproduces $T_e(X)$. 
The fitting parameters are shown in Fig.~\ref{fig5} (b) and (c).

$T_0$ does not depend on the laser intensity, which is consistent with the results in Fig.~\ref{fig3}.
$T_0$ is the dominant parameter at a deeper region, because $T_2$ is small. 
$T_1$, which corresponds to $T_e(X=0)$, increases monotonically as the function of laser intensity. 
$T_1$ and $T_2$ are close to each other at $5\times10^{12}$ and $7\times10^{12}$ W/cm$^2$.
This similarity indicates that the multiphoton absorption is the dominant excitation process at such intensities.
However, above $1\times10^{13}$  W/cm$^2$, the deviation between $T_1$ and $T_1$ indicates that a change in the 
excitation process occurs at the surface, as shown in Fig.~\ref{fig4}.
The damping parameter $\tau_1$ decreases as the laser intensity increases, which 
corresponds to the steep decrease of $T_e$ around the surface.

\subsection{3C-SiC}
\begin{figure} 
\includegraphics[width=85mm]{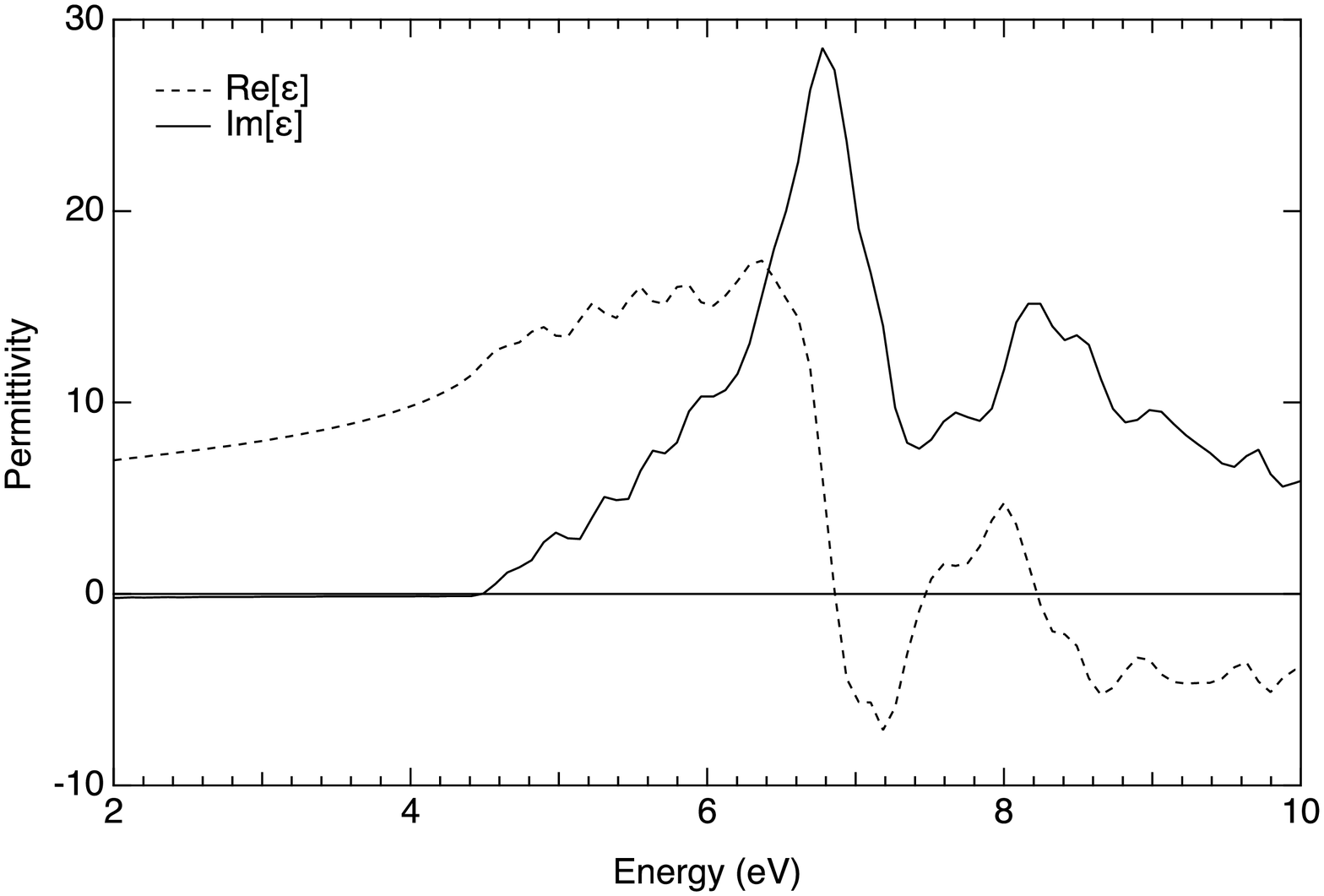} 
\caption{\label{fig6} 
Dielectric function of 3C-SiC.}
\end{figure}

\begin{figure} 
\includegraphics[width=85mm]{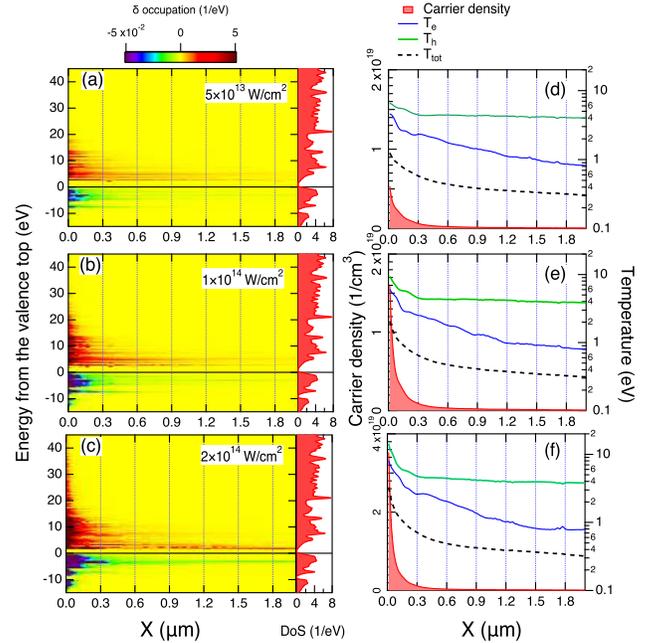} 
\caption{\label{fig7} 
(a)--(c) The change of electron occupation in 3C-SiC as a function of the position from the surface together with the DoS.
(d)--(f) The carrier density, quasi-temperatures ($T_e$ and $T_h$), and temperature ($T_{tot}$). }
\end{figure} 

\begin{figure} 
\includegraphics[width=85mm]{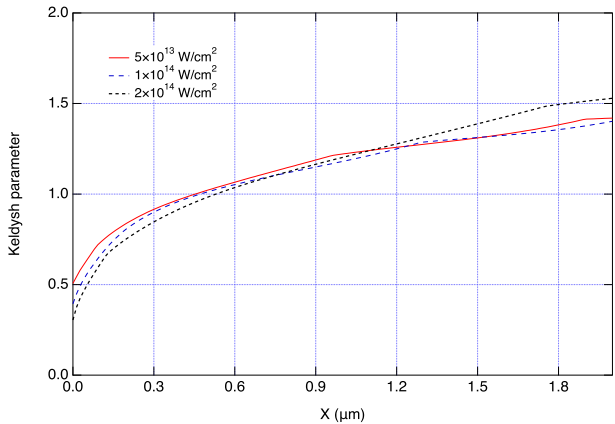} 
\caption{\label{fig8} 
Position dependence of the Keldysh parameter in 3C-SiC.}
\end{figure} 
SiC is attracting interest as a foundation base in the next generation, 
because of its wide band gap ($2\sim3$ eV),  temperature resistance, 
good thermal conductivity, as well as impact resistance \cite{Persson96,Morkoc94, Matsunami98,Son00}.
However, for its hardness and chemical and mechanical stability, SiC is difficult to process.
The processing by the femtosecond laser pulse enables the processing of SiC by the nonlinear processes \cite{Choi16}.

Figure~\ref{fig6} shows the dielectric function of 3C-SiC with LDA potential.
From the imaginary part of the dielectric function (Im$[\varepsilon]$), the optical band gap is 4.2 eV.

In the macroscopic calculation, we discretized a cubic cell, including 4 carbon and 4 silicon atoms, into $20\times 20\times 20$ grids.
We used $8\times8\times8$ K-sampling grids for the Bloch phase space.
Eighty-four conduction bands were prepared to calculate the excited electron distribution.

Figure \ref{fig7} (a)--(c) shows the position-dependent electron-hole distribution at three different laser intensities.
The hole density has a peak at -3 eV and the electron density has a peak at 3 eV at a deeper region, whose energy difference
(6 eV) corresponds to four-photon absorption.

Although the quasi-temperatures of silicon show a smooth decrease with increasing $X$,
 3C-SiC shows a stepwise structure in Fig.~\ref{fig6} (d)--(f).
 The abrupt change in $T_e$ and $T_h$ corresponds to the change of the number of contributing bands.
 The stepwise structure indicates that the band structure affects the electron-hole distribution and temperature.
 However, the carrier density and $T_{tot}$ show monotonic behavior as $X$ increases.
At all intensities, $T_h$ becomes position-independent at $X > 0.3~\mu m$, which corresponds to the behavior of the 
Keldysh parameter, as shown in Fig.~\ref{fig8}.

While $T_e$ is much higher than $T_h$ in silicon, $T_h$ of 3C-SiC becomes much higher than $T_e$.
Although $T_h$ at a deeper region is 4 eV at all laser intensities, which corresponds to hole creation at same energy point,
$T_e$  decreases as $X$ increases. 
From Fig.~\ref{fig7} (a)-(c), the electron at 5 eV survives even at a relatively deeper position (weak field intensity).
These results indicate that the quasi-temperatures depend strongly on the material, in other words, band structure.
  

\section{Summary}
In summary, we studied the spatial dependence of the quasi-temperatures in silicon and 3C-SiC by employing the 
time-dependent Kohn--Sham equation and Maxwell equation.
We found that the quasi-temperature and the electron-hole distribution depend on the excitation process.
While the quasi-temperatures do not show any spatial dependence where multiphoton absorption is dominant,
they increase at the surface where the tunneling excitation process is important.
Although we found that the carrier quasi-temperature can be approximated by a simple double exponential function in silicon,
3C-SiC shows a stepwise spatial distribution in quasi-temperatures.
These results indicate that the estimation of the quasi-temperatures from the quantum mechanical simulation is important 
to employ electron-lattice models, such as the two-temperature model, to understand the initial state of the laser processing and/or laser damage.

Our method treats the electron and electromagnetic field dynamics without any artificial parameters.
However, we should contain lattice dynamics induced by the electron excitation, which occurs in longer time scale, to simulate the 
laser processing.
We can understand the early stage of the laser processing by combining our method and classical molecular dynamics or fluid dynamics in the future.

\section*{Acknowledgement}
This work was supported by MEXT Quantum Leap Flagship Program (MEXT Q-LEAP) Grant Number JPMXS0118067246, 
JST-CREST under grant number JP-MJCR16N5, and by JSPS KAKENHI Japan Grant Numbers JP17H03525.
The numerical calculations were performed on the supercomputer SGI ICE X at 
the Japan Atomic Energy Agency (JAEA).
We would like to thank Editage (www.editage.com) for English language editing.

\bibliography{LaserProc2.bib}

\end{document}